\RequirePackage{lineno}
\documentclass[aps,prl,twocolumn,superscriptaddress,amsmath,amssymb,preprintnumbers]{revtex4-2}
\def\revtex{true}

\usepackage{etoolbox} 
\usepackage{graphicx} 
\usepackage{epstopdf} 
\usepackage{dcolumn} 
\usepackage{xcolor} 
\usepackage{amsmath,amssymb} 
\usepackage{hyperref} 
\usepackage[utf8]{inputenc} 
\usepackage[english]{babel} 
\usepackage{ifthen} 
\usepackage{blindtext} 
\usepackage{orcidlink} 
\usepackage{datetime2} 
\ifthenelse{\isundefined{\revtex}}{\usepackage[displaymath, mathlines]{lineno}}{} 

\graphicspath{{figures/}} 

\newboolean{articletitles}
\setboolean{articletitles}{true} 

\RequirePackage{xspace}
\RequirePackage{relsize}

\def\belletwo{\mbox{Belle~II}\xspace}

\def\babar{\mbox{\slshape B\kern-0.1em{\smaller A}\kern-0.1em
    B\kern-0.1em{\smaller A\kern-0.2em R}}\xspace}

\mathchardef\PDelta="7101
\mathchardef\PXi="7104
\mathchardef\PLambda="7103
\mathchardef\PSigma="7106
\mathchardef\POmega="710A
\mathchardef\PUpsilon="7107
                 
\def\PB      {\ensuremath{B}\xspace}                 
                 
\def\PD      {\ensuremath{D}\xspace}

\def\PK      {\ensuremath{K}\xspace}

\def\kaon  {\ensuremath{\PK}\xspace}
\def\Kbar  {\kern 0.2em\overline{\kern -0.2em \PK}{}\xspace}

\def\Kz    {\ensuremath{\kaon^0}\xspace}
\def\Kzb   {\ensuremath{\Kbar^0}\xspace}
\def\KzKzb {\ensuremath{\Kz \kern -0.16em \Kzb}\xspace}
\def\Kp    {\ensuremath{\kaon^+}\xspace}
\def\Km    {\ensuremath{\kaon^-}\xspace}

\def\KpKm  {\ensuremath{\Kp \kern -0.16em \Km}\xspace}

\def\D       {\ensuremath{\PD}\xspace}
\def\Dbar    {\kern 0.2em\overline{\kern -0.2em \PD}{}\xspace}

\def\Dz      {\ensuremath{\D^0}\xspace}
\def\Dzb     {\ensuremath{\Dbar^0}\xspace}
\def\DzDzb   {\ensuremath{\Dz {\kern -0.16em \Dzb}}\xspace}
\def\Dp      {\ensuremath{\D^+}\xspace}
\def\Dm      {\ensuremath{\D^-}\xspace}

\def\DpDm    {\ensuremath{\Dp {\kern -0.16em \Dm}}\xspace}

\def\Bbar    {\ensuremath{\kern 0.18em\overline{\kern -0.18em \PB}{}}\xspace}

\def\Y#1S{\ensuremath{\PUpsilon{(#1S)}}\xspace}

\def\Lbar {\ensuremath{\kern 0.1em\overline{\kern -0.1em\PLambda}}\xspace}

\def\to                 {\ensuremath{\rightarrow}\xspace}

\newcommand{\tev}{\ensuremath{\mathrm{\,Te\kern -0.1em V}}\xspace}
\newcommand{\gev}{\ensuremath{\mathrm{\,Ge\kern -0.1em V}}\xspace}
\newcommand{\mev}{\ensuremath{\mathrm{\,Me\kern -0.1em V}}\xspace}
\newcommand{\kev}{\ensuremath{\mathrm{\,ke\kern -0.1em V}}\xspace}
\newcommand{\ev}{\ensuremath{\mathrm{\,e\kern -0.1em V}}\xspace}
\newcommand{\gevc}{\ensuremath{{\mathrm{\,Ge\kern -0.1em V\!/}c}}\xspace}
\newcommand{\mevc}{\ensuremath{{\mathrm{\,Me\kern -0.1em V\!/}c}}\xspace}
\newcommand{\gevcc}{\ensuremath{{\mathrm{\,Ge\kern -0.1em V\!/}c^2}}\xspace}
\newcommand{\gevgevcccc}{\ensuremath{{\mathrm{\,Ge\kern -0.1em V^2\!/}c^4}}\xspace}
\newcommand{\mevcc}{\ensuremath{{\mathrm{\,Me\kern -0.1em V\!/}c^2}}\xspace}

\def\gsim{{~\raise.15em\hbox{$>$}\kern-.85em
          \lower.35em\hbox{$\sim$}~}\xspace}
\def\lsim{{~\raise.15em\hbox{$<$}\kern-.85em
          \lower.35em\hbox{$\sim$}~}\xspace}

\begin{document}

\title{Search for axion-like particles decaying to two photons at Belle~II}

  \author{M.~Abumusabh\,\orcidlink{0009-0004-1031-5425}} 
  \author{I.~Adachi\,\orcidlink{0000-0003-2287-0173}} 
  \author{A.~Aggarwal\,\orcidlink{0000-0002-5623-3896}} 
  \author{H.~Ahmed\,\orcidlink{0000-0003-3976-7498}} 
  \author{Y.~Ahn\,\orcidlink{0000-0001-6820-0576}} 
  \author{H.~Aihara\,\orcidlink{0000-0002-1907-5964}} 
  \author{M.~Akdag\,\orcidlink{0009-0004-3728-1077}} 
  \author{N.~Akopov\,\orcidlink{0000-0002-4425-2096}} 
  \author{S.~Alghamdi\,\orcidlink{0000-0001-7609-112X}} 
  \author{M.~Alhakami\,\orcidlink{0000-0002-2234-8628}} 
% ...
\collaboration{The Belle II Collaboration}

\begin{abstract}
Axion-like particles (ALPs) are predicted in many extensions of the Standard Model and provide a well-motivated portal between visible and hidden sectors through their coupling to photons.
We search for ALPs produced in the process $e^{+}e^{-}\to\gamma a$, $a\to\gamma\gamma$, using a data sample corresponding to an integrated luminosity of $408\,\mathrm{fb}^{-1}$ recorded by the Belle~II detector at the SuperKEKB $e^{+}e^{-}$ collider. 
Events containing three photons are used to reconstruct the ALP as a narrow peak in the di-photon invariant mass spectrum over the range $0.17 < m_{a} < 9.80~\mathrm{GeV}/c^{2}$. 
No significant excess above background is observed.
We set 95\% confidence level upper limits on the production cross section and on the ALP–photon coupling $g_{a\gamma\gamma}$, reaching sensitivities at the level of $10^{-4}~\mathrm{GeV}^{-1}$. 
The limits are the most restrictive to date over nearly the entire mass range $0.17 < m_{a} < 5.00~\mathrm{GeV}/c^{2}$, and improve upon previous results by up to a factor 9.
\end{abstract}

\maketitle

Axion-like particles (ALPs) appear in many extensions of the standard model (SM)~\cite{Jaeckel:2010ni}. 
They share the same quantum numbers as the QCD axion~\cite{PhysRevLett.38.1440}, but their mass and couplings are not linked. 
Light ALPs, around the MeV scale and below, are relevant for astrophysics and cosmology and may act as cold dark matter~\cite{PaolaArias_2012}. 
Heavier states can mediate interactions between dark matter and SM particles~\cite{Dolan:2017osp}.
ALPs that couple to the neutral electroweak sectors via $\gamma$ and $Z$-bosons are much less constrained than those with couplings to fermions or gluons. 
The latter often induce rare flavor-changing processes, which are strongly constrained~\cite {Dolan:2014ska}. 
In this work, we search for an ALP $a$ with mass $m_a$ that couples dominantly to photons with coupling strength $g_{a\gamma\gamma}$. 
We assume the coupling to $\gamma Z$ is negligible, so that the branching fraction $a \to \gamma\gamma$ is unity~\cite{Dolan:2017osp}.

In the $\mathrm{MeV}/c^{2}$ to $\mathrm{GeV}/c^{2}$ mass range, the strongest limits for photon-coupled ALPs come from a variety of production modes at different accelerator-based experiments. 
Light ALPs are constrained by reinterpretations of \hbox{$e^{+}e^{-}\!\to\!\gamma+\text{invisible}$} and beam-dump experiments~\cite{Dolan:2017osp,PhysRevD.38.3375,Blumlein1991,Blumlein1992}, and by searches at fixed-target experiments~\cite{NA64:2020qwq} and at forward facilities~\cite{FASER:2024bbl}.
Intermediate masses are constrained by reinterpretations of $e^{+}e^{-}\!\to\!\gamma\gamma$~\cite{PhysRevLett.118.171801,Abbiendi2003} and coherent Primakoff production on nuclei~\cite{PhysRevLett.123.071801,PrimEx:2010fvg}, and by searches for $e^{+}e^{-}\!\to\!\gamma a$~\cite{PhysRevLett.125.161806}, and for $J/\psi\to\gamma a$ decays~\cite{2023137698, PhysRevD.110.L031101}.
Heavier ALPs are constrained by photon-photon fusion in peripheral heavy-ion collisions~\cite{ATLAS:2020hii,2019134826}.

In this letter, we search for ALPs produced in the radiative process $e^+e^-\to\gamma a$ using data collected by the \belletwo experiment at the SuperKEKB collider.
We use events with three photons to search for an ALP signal as a narrow peak in the reconstructed di-photon mass distribution resulting from $a\to\gamma\gamma$ decays.
Within our search range, the ALP width is negligible relative to the experimental resolution, and the ALP decays promptly. 
The probed ALP masses span the range $0.17 < m_a < 9.80$\,$\mathrm{GeV}/c^{2}$.
The lower limit is determined by the irreducible $\pi^0$ background, whereas the upper limit is determined by the event selection, which is optimized to suppress the rapidly increasing background from SM processes and to ensure stable signal extraction.

The \belletwo experiment~\cite{Abe:2010gxa} operates at the SuperKEKB electron–positron collider~\cite{Akai:2018mbz} located at KEK in Tsukuba, Japan. 
The beam energies are 7~GeV for electrons and 4~GeV for positrons, corresponding to a boost of $\beta\gamma = 0.28$ of the center-of-mass (c.m.) frame relative to the laboratory frame. 
The data used in this analysis were recorded at a center-of-mass energy $\sqrt{s}$ corresponding to the mass of the $\Upsilon(4S)$ resonance, as well as at an energy approximately 60~MeV below this resonance. 
The total integrated luminosity amounts to $408.1 \pm 1.7$\,fb$^{-1}$~\cite{luminosity}, of which 42.6\,fb$^{-1}$ were collected below the $\Upsilon(4S)$ resonance.

The \belletwo detector consists of a variety of sub-detectors surrounding the interaction point~(IP) in a cylindrical manner.  
The trajectories of charged particles~(tracks) are reconstructed by a combination of a two-layer silicon pixel detector, a four-layer silicon strip detector, and a central drift chamber.  
The tracking detectors cover an angular region of $17^{\circ} < \theta_{\mathrm{polar}} < 150^{\circ}$ and are surrounded by time-of-propagation~\cite{Kotchetkov:2018qzw} and aerogel ring-imaging Cherenkov detectors used for particle identification.
Photons are reconstructed by an electromagnetic calorimeter~(ECL) that also serves in the identification of electrons, covering $12^{\circ} < \theta_{\mathrm{polar}} < 155^{\circ}$.
The ECL is surrounded by a $1.5\,\mathrm{T}$ superconducting solenoid.  
The outermost subdetector is a $K^{0}_{L}$ and muon detector, which is installed in the iron flux return of the solenoid.  
The longitudinal direction, the transverse plane, and the polar angle $\theta_{\mathrm{polar}}$ are defined with respect to the detector’s solenoidal axis in the direction of the electron beam.  
In the following, quantities are defined in the laboratory frame unless specified otherwise.  

Simulated events are used to optimize the event selection and determine signal efficiencies and resolutions. 
Signal samples are generated with \textsc{MadGraph5@NLO}~\cite{Alwall:2014hca}, including initial-state radiation (ISR)~\cite{Frixione:2021zdp}, for several hypotheses of $m_a$ in step sizes matching approximately half the signal mass resolution. 
To account for variations in efficiency due to varying beam-background conditions, simulated signal events are overlaid with beam-induced backgrounds sampled from data. 
An approximately linear dependence of the signal reconstruction efficiency on the background level is observed, and the efficiency at dataset's luminosity-weighted average background level is determinedined from a linear fit.

The SM background processes $e^{+}e^{-} \to e^{+}e^{-}(\gamma)$ and $e^{+}e^{-} \to \gamma\gamma(\gamma)$ are simulated with \textsc{Babayaga@NLO}~\cite{Balossini:2008xr}. 
Here, the notation $(\gamma)$ indicates that any number of additional photons, including none, may be produced by the generator.
Background processes $e^{+}e^{-} \to h^0\gamma(\gamma)$, where $h^0$ is a pseudoscalar meson ($\pi^{0}$, $\eta$, $\eta'$), are simulated with \textsc{Phokhara9}~\cite{PhysRevD.97.016006}, including production through the radiative decays of the intermediate vector resonances $\rho$, $\omega$, and $\phi$. 
The same generators are used to compute the corresponding cross sections.
Other SM background processes were found to be negligible.

The detector response is simulated with \textsc{Geant4}~\cite{Agostinelli:2002hh}.
Event reconstruction and analysis are performed with the \belletwo analysis software framework~\cite{Kuhr:2018lps, basf2-zenodo}.
To avoid the experimenter’s bias, we examine the experimental data only after finalizing the analysis selection and fitting procedure. 

All selection criteria are optimized to maximize a figure of merit for a discovery with a significance of five standard deviations~\cite{Punzi:2003bu}. 
The following selections are applied to all ALP mass search windows, as the differences compared to a mass-dependent optimization are found to be small.

Photon candidates are reconstructed from clusters in the ECL without associated charged tracks. 
They are required to have an energy above $0.5~\mathrm{GeV}$, a polar angle $17^{\circ} < \theta_{\mathrm{polar}} < 150^{\circ}$, and a minimum distance in the transverse plane of $50~\mathrm{cm}$ to the nearest track. 
This distance is measured between the cluster centroid and the extrapolation of the track to the calorimeter surface.
The cluster timing must satisfy $|T_{\text{cluster}}| < 200~\mathrm{ns}$ with respect to the global event time.

Event candidates are formed from the three most energetic photons $\gamma_1$, $\gamma_2$ and $\gamma_3$ in each event, where $\gamma_1$ is the highest energetic photon in the event. 
The ALP candidate is reconstructed from any two photons, yielding three ALP candidates per event.

The helicity angle $\lambda$ of the ALP is defined as the angle between the momentum of one decay photon and the momentum of the initial $e^+e^-$ system, both evaluated in the ALP rest frame. 
For signal events, the helicity angle of the ALP is expected to be uniformly distributed, while scattering processes such as $e^{+}e^{-} \to \gamma\gamma(\gamma)$ exhibit a strong preference for $\cos\lambda = \pm 1$.
Each ALP candidate is required to satisfy $|\cos\lambda| < 0.74$.
The helicity angle selection reduces the number of candidates per event from three to an average of $1.24$.

Furthermore, the clusters in each event are required to be consistent in time with one another and the global ECL time.
Their reconstructed mass is required to satisfy $0.80\,\sqrt{s} < M_{\gamma_1\gamma_2\gamma_3} < 1.05\,\sqrt{s}$.
Events with two or more charged tracks originating from the IP are rejected to reduce backgrounds from radiative Bhabha scattering.
Events with one charged track originating from the IP are allowed in the selection due to the existence of beam-background induced tracks.
We require the sum of the energies of all clusters in the ECL in the polar-angle region $22^{\circ} < \theta_{\mathrm{polar}} < 128^{\circ}$ to be between $4~\mathrm{GeV}$ and $13~\mathrm{GeV}$ to reject events with very high beam-background energy depositions in the ECL.

A kinematic fit is applied to improve the invariant mass resolution of the diphoton system by constraining the sum of the three-photon final state four-momentum to the known initial state four-momentum. 
This results in narrower $M_{\gamma\gamma}$ distributions and a weaker dependence of the signal mass resolution on the photon energy.
The fit also aligns the reconstructed peak position more closely with the generated ALP mass, eliminating the need for additional bias corrections.

To suppress backgrounds from photon conversions occurring outside the tracking detector volume, events are rejected if the angular differences for photon pairs satisfy $\left|\Delta\theta_{1,3}\right| < 0.014~\mathrm{rad}$ and $\left|\Delta\phi_{1,3}\right| < 0.4~\mathrm{rad}$, or $\left|\Delta\theta_{2,3}\right| < 0.014~\mathrm{rad}$ and $\left|\Delta\phi_{2,3}\right| < 0.4~\mathrm{rad}$.
A $\pi^{0}$ veto is applied by requiring $M_{\gamma_i\gamma_j} > 0.16~\mathrm{GeV}/c^{2}$ for all photon pairs $\gamma_i\gamma_j$ constructed from the four highest energetic photons.

An event classifier based on a feedforward neural network~\cite{Rumelhart1986} is used to separate signal from the largest $e^+e^-\to \gamma\gamma(\gamma)$ background. 
The network takes 19 input features that describe the event kinematics and topology, including the invariant mass and the absolute momentum in the center-of-mass frame of the three-photon system, individual photon energies, the energy product of the photon pairs, the differences in polar and azimuthal angles, the three-dimensional opening angle, the event sphericity, and the ratio of second to zeroth Fox–Wolfram moments.
While no single input feature separates signal from background across all ALP mass hypotheses, the network exploits their combined discriminating power, motivating a single classifier trained uniformly over all mass hypotheses.
The network is trained on a class-balanced sample consisting of simulated events generated under different beam-background conditions and signal events with a uniform distribution in mass. 

In addition, we retain events that satisfy at least one of two trigger conditions. 
Either the total energy deposition in the ECL in the polar-angle region $22^{\circ} < \theta_{\text{polar}} < 128^{\circ}$ exceeds $1~\mathrm{GeV}$ and the event passes a veto of two back-to-back high energy clusters, or the event contains two back-to-back clusters in the ECL and no nearby charged tracks.
In both cases, clusters are required to have energies in the c.m. frame above $3~\mathrm{GeV}$, with at least one cluster above $4.5~\mathrm{GeV}$.
The trigger efficiency is close to 100\% for all ALP masses passing the previously described selection due to the complement of the aforementioned trigger decisions.

The overall signal selection efficiency is typically $25$-$35\%$ for masses up to about $7~\mathrm{GeV}/c^{2}$ and decreases approximately linearly to about $10\%$ at the highest masses.
The resulting $M_{\gamma\gamma}$ distributions are shown in Fig.~\ref{fig:datamc} together with the stacked contributions from simulated SM background samples normalized to the recorded integrated luminosity.
From simulation we conclude that more than $99\%$ of the remaining events originate from the processes $e^{+}e^{-} \to \gamma\gamma(\gamma)$.

\begin{figure}[ht]
    \centering
    \includegraphics[width=0.49\textwidth]{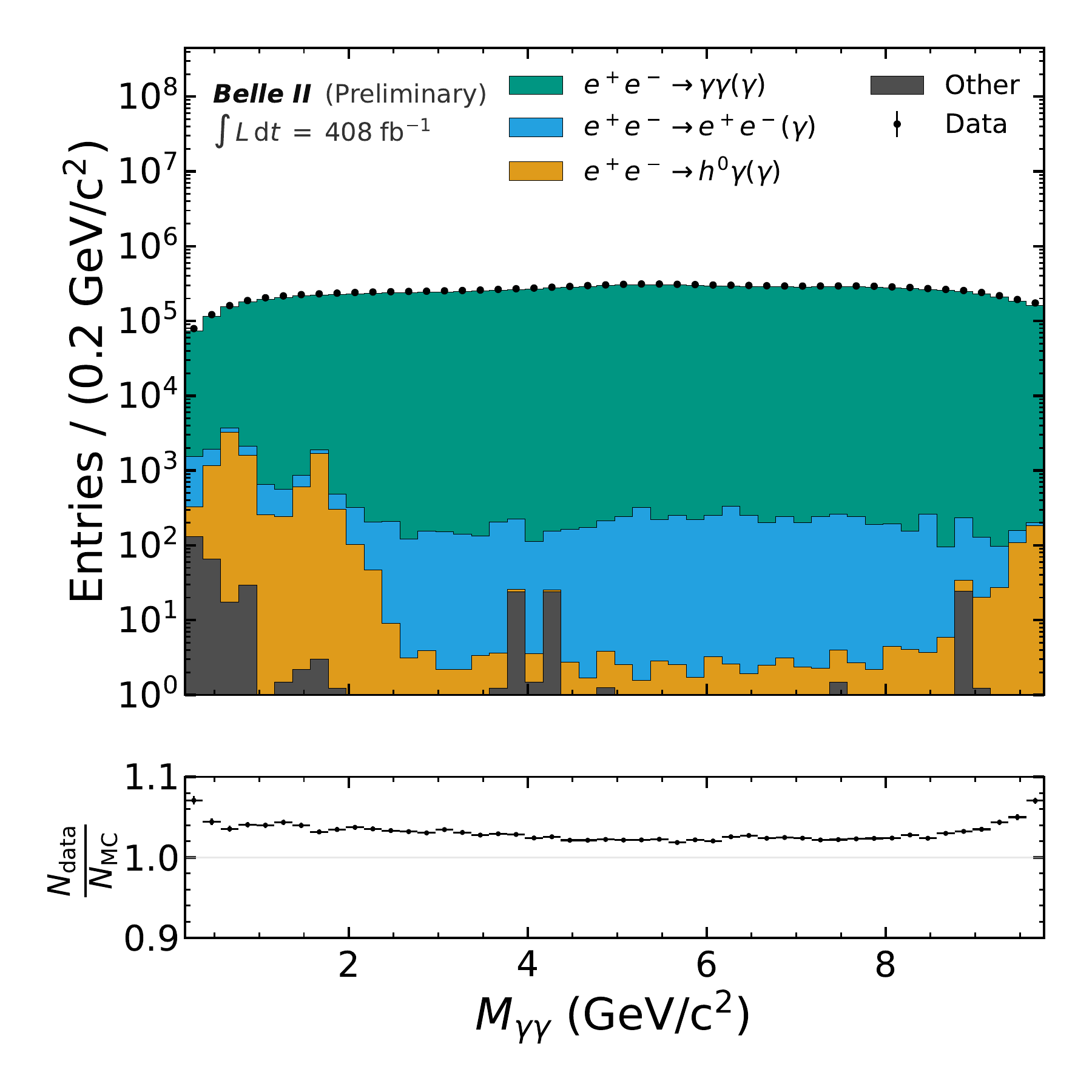}
    \caption{
        Distribution of $M_{\gamma\gamma}$ together with the stacked contributions from the various simulated SM background samples. Simulation is normalized to a luminosity of $408\,\mathrm{fb}^{-1}$.
    }\label{fig:datamc}
\end{figure}

Signal hypotheses are formed in the invariant di-photon mass distribution $M_{\gamma\gamma}$ and are tested by extended unbinned maximum-likelihood fits.
The signal probability density function (PDF) consists of two components: a peaking contribution from correctly reconstructed signal photons and a combinatorial contribution from incorrect combinations of a signal photon and a recoil photon.
The peaking component is modelled with a double-sided Crystal Ball (DSCB) function~\cite{Gaiser:Phd, Skwarnicki:1986xj,Eschle:2019jmu}, whose parameters are determined from fits to simulated $M_{\gamma\gamma}$ distributions for all scanned $m_a$ samples. 
The resolution $\sigma^{\text{sig}}_{\text{DSCB}}$ after kinematic fit increases smoothly from about $10~\mathrm{MeV}/c^{2}$ for the smallest to about $16~\mathrm{MeV}/c^{2}$ for the largest ALP masses.  
The combinatorial signal component is described by an additional first-order polynomial, with its normalization relative to the correctly reconstructed component fixed from simulation.
This normalization ranges from $0\%$ at very low and very high ALP masses to a maximum of $\sim30\%$ near $m_a=7~\mathrm{GeV}/c^{2}$, beyond which it falls off rapidly.
In the fits to data, the mass-dependent signal parameters are fixed to the values obtained from simulation.

The background is described by polynomial functions whose order is determined from simulation to provide the minimal adequate description: fourth order for $0.17 < m_{a} < 1.00~\mathrm{GeV}/c^{2}$ and $8.50 < m_{a} < 9.80~\mathrm{GeV}/c^{2}$, and second order elsewhere.
All polynomial coefficients are left free in the fits. 
We exclude ALP candidates with $0.45 < m_{a} < 0.63~\mathrm{GeV}/c^{2}$ to remove peaking backgrounds from $\eta \to \gamma\gamma$ decays, and $0.72 < m_{a} < 0.81~\mathrm{GeV}/c^{2}$ to suppress contributions from $\omega \to \pi^{0}\gamma$ decays. 
The contribution from $\eta' \to \gamma\gamma$ decays is found to be negligible.

We extract the signal cross-section as a function of $m_{a}$ by performing a series of independent fits~\cite{Eschle:2019jmu} in $\pm20\sigma^{\text{sig}}_{\text{DSCB}}$ signal mass windows in steps of $5~\mathrm{MeV}$. 
This step size is always smaller than half of the signal mass resolution.

We evaluate systematic uncertainties affecting the photon detection efficiency, photon resolutions, selection efficiency, ISR modeling, trigger efficiency, differences of beam-background dependent simulation, luminosity and signal shape.
The differences between data and simulation for the photon reconstruction efficiency, the energy and the spatial resolutions are measured using radiative muon-pair events.
Event efficiency differences of 0.5--2\% are corrected for, and an uncertainty of approximately 2.5\% per event is assigned due to this correction, showing only small variation with \(m_{a}\).
The energy resolution is found to be $17\%$ worse in data than in simulation, while the spatial resolutions are found to be in agreement.
Integrated over the entire photon energy spectrum, the relative energy resolution is estimated to be approximately $1.4\%$ in data, while in simulation we find $1.2\%$.
To estimate the impact on the signal shape, we adjust the photon energy for the observed difference during the signal reconstruction.
The signal resolution $\sigma^{\text{sig}}_{\text{DSCB}}$ after kinematic fit shows agreement within statistical uncertainty with and without the adjusted energy.
No additional systematic uncertainty is assigned for differences in photon energy resolution and the uncertainty in the signal shape arises from the limited number of simulated signal events.
The uncertainty of the energy bias correction, determined using symmetric $\pi^0\to\gamma\gamma$ and $\eta\to\gamma\gamma$ decays, is also found to be negligible. 

The trigger efficiency is measured directly on the selected data and found to agree with simulation, being close to 100\%.
We estimate the uncertainty of the trigger efficiency determination from the statistical precision to be $0.01\%$ across all ALP mass scan points.

The uncertainty due to the interpolation of the signal efficiency for different beam-background conditions is estimated to be about $0.2\%$, independent of the ALP mass.

A relative systematic uncertainty of 0.42\% is assigned to the integrated luminosity. 
The impact of ISR modelling is evaluated by comparing signal samples generated with \textsc{WHIZARD}~\cite{Kilian:2007gr, Moretti:2001zz} and \textsc{MadGraph@NLO}, and the observed difference in the signal efficiency is taken as the systematic uncertainty.
This systematic uncertainty ranges from approximately $7\%$ at low ALP masses and decreases to approximately $1\%$ at the highest masses.
The full difference between data and simulation normalization is also treated as a systematic contribution to the selection efficiency with a relative value of $4\%$. 
The total systematic uncertainty of the signal efficiency is obtained by adding all individual contributions in quadrature.
It ranges from about 8\% at low ALP masses to about 4\% at high masses.
For low ALP masses, the uncertainty is dominated by ISR modelling, while for high masses it is dominated by the selection efficiency. 

The systematic uncertainties due to the signal efficiency, the luminosity, and the signal shape are included as Gaussian nuisance parameters with a width equal to the systematic uncertainty.
A systematic uncertainty due to the choice of the background shape polynomial order is found to be negligible.

The local significance $S$ of the signal for a given mass hypothesis is defined as $S = \sqrt{2(\ln\mathcal{L} - \ln\mathcal{L}_{\text{bkg}})}$, where $\mathcal{L}$ is the maximum likelihood from the signal-plus-background fit and $\mathcal{L}_{\text{bkg}}$ is the maximum likelihood for the background-only hypothesis. 
The largest local significance, including systematic uncertainties, is $3.3\sigma$, found near $m_{a} = 0.22~\mathrm{GeV}/c^{2}$.
Taking into account the look-elsewhere effect~\cite{Gross:2010qma}, this excess corresponds to a global significance of $1.4\sigma$.

We compute the upper limits at 95\% confidence level (C.L.) on the cross section $\sigma_{a}=\sigma(e^+e^-\to\gamma a, a\to\gamma\gamma)$ as a function of $m_{a}$ using a two-sided frequentist profile-likelihood method~\cite{Cowan2011}.
The expected limit under the background-only hypothesis and the observed upper limits on $\sigma_{a}$ are shown in Fig.~\ref{fig:brazil}. 

To convert the limits on the cross-section to the coupling $g_{a\gamma\gamma}$, we compare the fitted limits with the simulated cross-sections for a fixed coupling.
This calculation does not account for possible energy dependence of $\alpha_{\text{QED}}$ or $g_{a\gamma\gamma}$~\cite{Bauer:2017ris}.
The observed upper limits on the ALP–photon coupling $g_{a\gamma\gamma}$, together with existing constraints from previous experiments, are shown in Fig.~\ref{fig:limits}.

\begin{figure}[ht]
    \centering
    \includegraphics[width=0.49\textwidth]{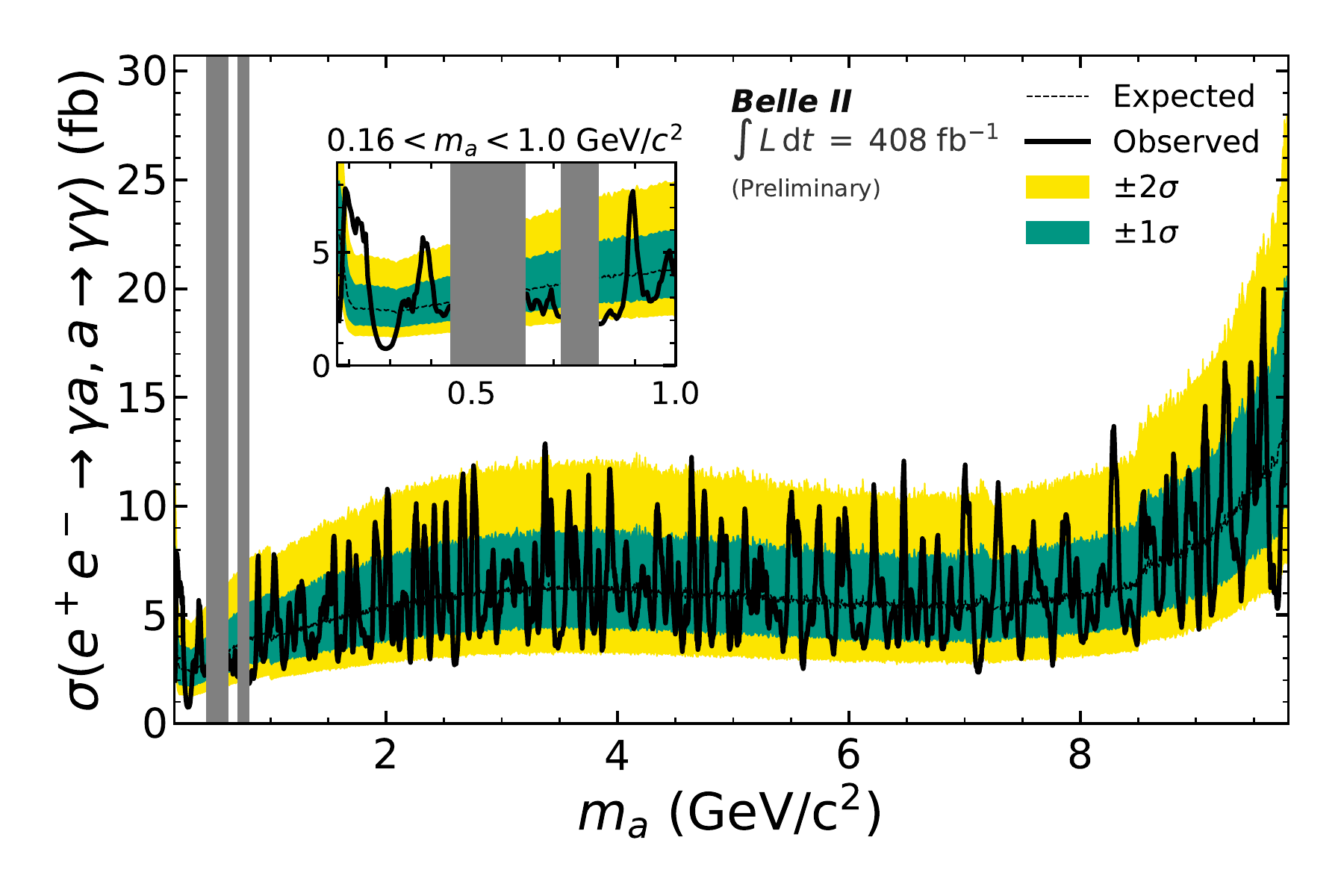}
    \caption{
        Expected and observed upper limits (95\% C.L.) on
        the cross section $\sigma(e^+e^-\to\gamma a, a\to\gamma\gamma)$ as a function of the ALP mass $m_a$. 
        The regions corresponding to the vetoed $\omega$ and $\eta$ masses are indicated in gray.
    }\label{fig:brazil}
\end{figure}

\begin{figure}[ht]
    \centering
    \includegraphics[width=0.49\textwidth]{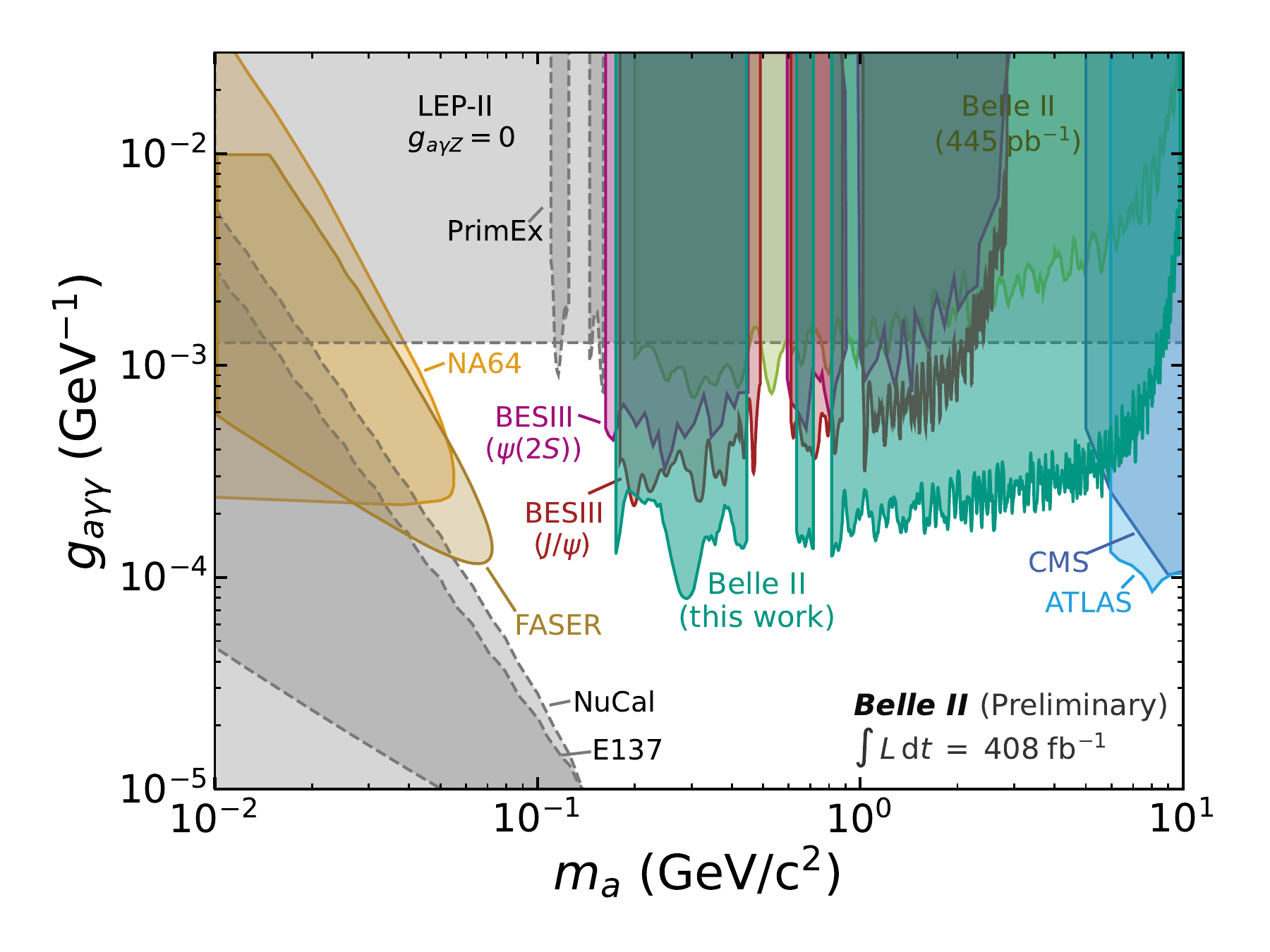}
    \caption{
        Exclusion region (95\% C.L.) in the plane of the ALP-photon coupling $g_{a\gamma\gamma}$ and ALP mass $m_a$ from this work~(teal) and previous constraints from \belletwo~(\cite{PhysRevLett.125.161806}), BES~III~(\cite{2023137698, PhysRevD.110.L031101}), CMS~(\cite{2019134826}), ATLAS~(\cite{ATLAS:2020hii}), FASER~(\cite{FASER:2024bbl}), and NA64~(\cite{NA64:2020qwq}). Also shown are constraints derived from reinterpretations not performed by the experimental collaborations, namely E137~(\cite{Dolan:2017osp,PhysRevD.38.3375}), LEP-II~(\cite{PhysRevLett.118.171801,Abbiendi2003}), PrimEx~(\cite{PhysRevLett.123.071801,PrimEx:2010fvg}) and NuCal~(\cite{Dolan:2017osp,Blumlein1991,Blumlein1992}), indicated by gray shading with dashed outlines.
    }\label{fig:limits}
\end{figure}

In conclusion, we have searched for the process $e^{+}e^{-} \to \gamma a$, $a \to \gamma\gamma$ over a wide ALP mass range using Belle~II data. 
This analysis improves upon the previous Belle~II result~\cite{PhysRevLett.125.161806}. 
The main improvements are the use of kinematic fits, an event-classifier neural network, and a substantially larger data sample, which together lead to a significant increase in sensitivity.
No significant excess of events consistent with the signal hypothesis is observed, and we set the most restrictive upper limits to date on the ALP–photon coupling $g_{a\gamma\gamma}$ over nearly the entire mass range $0.17 < m_{a} < 5.00~\mathrm{GeV}/c^{2}$, except near $m_a=0.22~\mathrm{GeV}/c^{2}$, and improve upon previous results by up to a factor 9.

\begin{acknowledgments}
This work, based on data collected using the Belle II detector, which was built and commissioned prior to March 2019,
was supported by
Higher Education and Science Committee of the Republic of Armenia Grant No.~23LCG-1C011;
Australian Research Council and Research Grants
No.~DP200101792, 
No.~DP210101900, 
No.~DP210102831, 
No.~DE220100462, 
No.~LE210100098, 
and
No.~LE230100085; 
Austrian Federal Ministry of Education, Science and Research,
Austrian Science Fund (FWF) Grants
DOI:~10.55776/P34529,
DOI:~10.55776/J4731,
DOI:~10.55776/J4625,
DOI:~10.55776/M3153,
and
DOI:~10.55776/PAT1836324,
and
Horizon 2020 ERC Starting Grant No.~947006 ``InterLeptons'';
Natural Sciences and Engineering Research Council of Canada, Digital Research Alliance of Canada, and Canada Foundation for Innovation;
National Key R\&D Program of China under Contract No.~2024YFA1610503,
and
No.~2024YFA1610504
National Natural Science Foundation of China and Research Grants
No.~11575017,
No.~11761141009,
No.~11705209,
No.~11975076,
No.~12135005,
No.~12150004,
No.~12161141008,
No.~12405099,
No.~12475093,
and
No.~12175041,
and Shandong Provincial Natural Science Foundation Project~ZR2022JQ02;
the Czech Science Foundation Grant No. 22-18469S,  Regional funds of EU/MEYS: OPJAK
FORTE CZ.02.01.01/00/22\_008/0004632 
and
Charles University Grant Agency project No. 246122;
European Research Council, Seventh Framework PIEF-GA-2013-622527,
Horizon 2020 ERC-Advanced Grants No.~267104 and No.~884719,
Horizon 2020 ERC-Consolidator Grant No.~819127,
Horizon 2020 Marie Sklodowska-Curie Grant Agreement No.~700525 ``NIOBE''
and
No.~101026516,
and
Horizon Europe Marie Sklodowska-Curie Staff Exchange project JENNIFER3 Grant Agreement No.~101183137 (European grants);
L’Institut National de Physique Nucl\'eaire et de Physique des
Particules (IN2P3) du CNRS under Project Identification No.
CNRS-IN2P3-14-PP-033
and L’Agence Nationale de la Recherche (ANR) under Grant No. ANR-23-CE31-
0018 and ANR-25-CE31-1333 (France);
BMFTR, DFG, HGF, MPG, and AvH Foundation (Germany);
Department of Atomic Energy under Project Identification No.~RTI 4002,
Department of Science and Technology,
and
UPES SEED funding programs
No.~UPES/R\&D-SEED-INFRA/17052023/01 and
No.~UPES/R\&D-SOE/20062022/06 (India);
Israel Science Foundation Grant No.~2476/17,
U.S.-Israel Binational Science Foundation Grant No.~2016113, and
Israel Ministry of Science Grant No.~3-16543;
Istituto Nazionale di Fisica Nucleare and the Research Grants BELLE2,
and
the ICSC – Centro Nazionale di Ricerca in High Performance Computing, Big Data and Quantum Computing, funded by European Union – NextGenerationEU;
Japan Society for the Promotion of Science, Grant-in-Aid for Scientific Research Grants
No.~16H03993,
No.~16H06492,
No.~16K05323,
No.~17H01133,
No.~17H05405,
No.~18K03621,
No.~18H03710,
No.~18H05226,
No.~19H00682, 
No.~20H05850,
No.~20H05858,
No.~22H00144,
No.~22K14056,
No.~22K21347,
No.~23H05433,
No.~26220706,
No.~26400255,
and
No.~26H02056,
and
the Ministry of Education, Culture, Sports, Science, and Technology (MEXT) of Japan;  
National Research Foundation (NRF) of Korea Grants
No.~2021R1-F1A-1064008,
No.~2022R1-A2C-1003993,
No.~RS-2018-NR031074,
No.~RS-2021-NR060129,
No.~RS-2024-00354342,
No.~RS-2025-02219521,
No.~RS-2026-25471491,
No.~RS-2026-25480677,
and
No.~RS-2026-25486791,
Radiation Science Research Institute,
Foreign Large-Size Research Facility Application Supporting project,
the Global Science Experimental Data Hub Center, the Korea Institute of Science and
Technology Information (K26L1M2C3)
and
KREONET/GLORIAD;
Universiti Malaya RU grant, Akademi Sains Malaysia, and Ministry of Education Malaysia;
Frontiers of Science Program Contracts
No.~FOINS-296,
No.~CB-221329,
No.~CB-236394,
No.~CB-254409,
and
No.~CB-180023, and SEP-CINVESTAV Research Grant No.~237 (Mexico);
the Polish Ministry of Science and Higher Education and the National Science Center;
the Ministry of Science and Higher Education of the Russian Federation
and
the HSE University Basic Research Program, Moscow;
University of Tabuk Research Grants
No.~S-0256-1438 and No.~S-0280-1439 (Saudi Arabia);
Slovenian Research Agency and Research Grants
No.~J1-50010
and
No.~P1-0135;
Ikerbasque, Basque Foundation for Science,
State Agency for Research of the Spanish Ministry of Science and Innovation through Grant No. PID2022-136510NB-C33, Spain,
the Severo Ochoa project CEX2023-001292-S funded by MICIU/AEI, State Secretariat for
Telecommunications and Digital Infrastructure with reference
TSI-069100-2023-0012, State Agency for Research of the Spanish Ministry
of Science, Innovation and Universities through Grant No
PID2024-156645NB-C21;
The Knut and Alice Wallenberg Foundation (Sweden), Contracts No.~2021.0174, No.~2021.0299, and No.~2023.0315;
National Science and Technology Council,
and
Ministry of Education (Taiwan);
Thailand Center of Excellence in Physics;
TUBITAK ULAKBIM (Turkey);
National Research Foundation of Ukraine, Project No.~2020.02/0257,
and
Ministry of Education and Science of Ukraine;
the U.S. National Science Foundation and Research Grants
No.~PHY-1913789 
and
No.~PHY-2111604, 
and the U.S. Department of Energy and Research Awards
No.~DE-AC06-76RLO1830, 
No.~DE-SC0007983, 
No.~DE-SC0009824, 
No.~DE-SC0009973, 
No.~DE-SC0010007, 
No.~DE-SC0010073, 
No.~DE-SC0010118, 
No.~DE-SC0010504, 
No.~DE-SC0011784, 
No.~DE-SC0012704, 
No.~DE-SC0019230, 
No.~DE-SC0021616, 
No.~DE-SC0022350, 
No.~DE-SC0023470; 
and
the Vietnam Academy of Science and Technology (VAST) under Grant
No.~DL0000.05/26-27.

These acknowledgements are not to be interpreted as an endorsement of any statement made
by any of our institutes, funding agencies, governments, or their representatives.

We thank the SuperKEKB team for delivering high-luminosity collisions;
the KEK cryogenics group for the efficient operation of the detector solenoid magnet and IBBelle on site;
the KEK Computer Research Center for on-site computing support; the NII for SINET6 network support;
and the raw-data centers hosted by BNL, DESY, GridKa, IN2P3, INFN, 
and the University of Victoria.

\end{acknowledgments}

\textit{Data availability} ---
\ifdefvoid{\hepdata}{}{Numerical data corresponding to the results presented are available as HEPData.}
The full Belle II data are not publicly available.
The collaboration will consider requests for access to the data that support this article.

\bibliography{belle2-references,references}

\clearpage
\appendix
\title{Supplemental Material: Search for axion-like particles decaying to two photons at \belletwo}
\maketitle

\renewcommand{\thefigure}{S\arabic{figure}}
\setcounter{figure}{0}


\section*{Fit Result at Maximum Local Significance}

\begin{figure}[!h]
    \centering
    \includegraphics[width=0.49\textwidth]{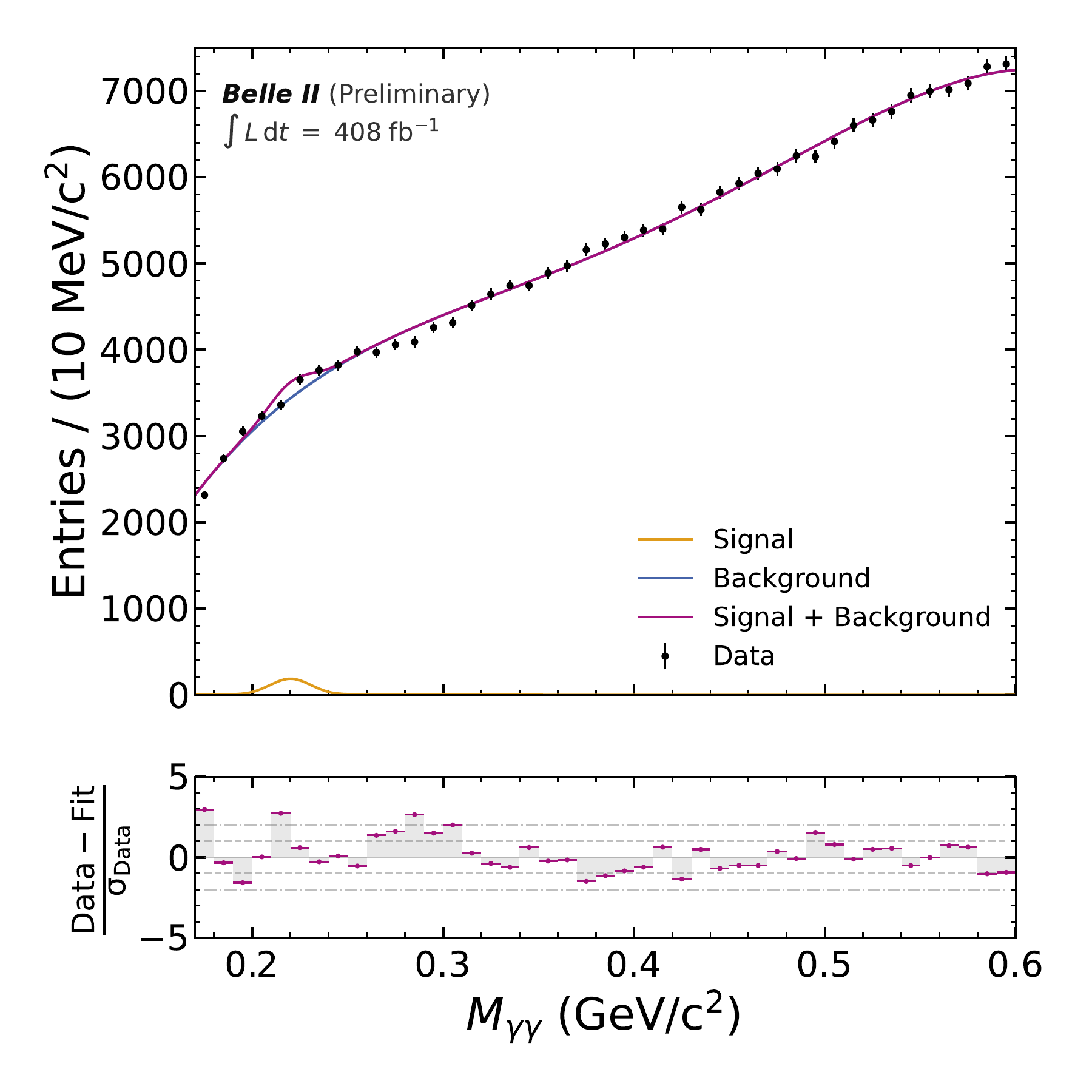}
    \caption{
        Unbinned maximum likelihood fit result for an ALP mass hypothesis of $m_a=0.22~\mathrm{GeV}/c^{2}$. Top: data distribution (black points), fitted signal (yellow), background (blue), and total (purple) PDFs. Bottom: pull between the data and the fitted total PDF.
    }\label{fig:significant}
\end{figure}

Fig.~\ref{fig:significant} shows the fit result for the ALP mass hypothesis $m_a=0.22~\mathrm{GeV}/c^{2}$, corresponding to the highest local (global) significance $S$ of $1.4\sigma$ ($3.3\sigma$) observed in the ALP mass scan.

\clearpage

\end{document}